\documentclass{aastex}
\usepackage{psfrag}
\usepackage{rotating}
\usepackage{microtype}
\usepackage{graphicx,shapepar}
\usepackage{lscape}

\setkeys{Gin}{width=\linewidth, height=0.9\textheight, keepaspectratio=true}

\newcommand{\kms}{\ensuremath{\mathrm{km\ s}^{-1}}}

\newcommand{\nii}{[\ion{N}{2}]}

\newcommand\NIIlam{[\ion{N}{2}]\,6584\,\AA\@}

\newcommand{\teff}{\ensuremath{T_\mathrm{eff}}}

\newcommand{\vheli}{\ensuremath{V_\mathrm{hel}}}

\newcommand{\vsys}{\ensuremath{V_\mathrm{sys}}}
\newcommand{\vexp}{\ensuremath{V_\mathrm{exp}}}
\newcommand{\grados}{$\,^\mathrm{o}\,$}

\title{Morpho-kinematic analysis of the point-symmetric, bipolar planetary nebulae Hb 5 and K 3-17, a pathway to poly-polarity.}

\author{J. A. L\'opez,
  Ma.~T. Garc\'{\i}a-D\'{\i}az,
   W. Steffen,  H. Riesgo, \& M. G. Richer
\affil{Instituto de Astronom\'{\i}a, Universidad
  Nacional Aut\'onoma de M\'exico} Campus
  Ensenada, Apdo. Postal 22860, Ensenada, B. C., M\'exico}
\affil{jal,tere,wsteffen,hriesgo,richer@astrosen.unam.mx}

\shorttitle{Dissecting  Hb~5 and K 3-17}
\shortauthors{L\'opez et al.}

\begin{abstract}

The kinematics of the bipolar planetary nebulae Hb~5 and K 3-17 are
investigated in detail by means of a comprehensive set of spatially
resolved high spectral resolution, long-slit spectra. Both objects
share particularly interesting characteristics, such as a complex
filamentary, rosette-type nucleus, axial point-symmetry and very fast
bipolar outflows.  The kinematic information of Hb~5 is combined with
{\it HST} imagery to construct a detailed 3D model of the nebula using the
code SHAPE.  The model shows that the large scale lobes are growing in
a non-homologous way. The filamentary loops in the core are proven to
actually be secondary lobes emerging from what appears to be a
randomly punctured, dense, gaseous core and the material that forms
the point symmetric structure flows within the lobes with a distinct
kinematic pattern and its interaction with the lobes has had a shaping 
effect on them. Hb~5 and K~3-17 may represent a class of fast
evolving planetary nebulae that will develop poly-polar
characteristics once the nebular core evolves and expands.

\end{abstract}


\begin{document}
\maketitle

\section{Introduction}
\label{sec:introduction}

Hb~5 and K~3-17 are akin in their peculiar characteristics and form a singular pair among the group of bipolar planetary nebulae (PNe). They both show tight waists with well developed bipolar lobes, complex nebular dense  cores composed of multiple, radial, filamentary loops that resemble a rosette, and structural point-symmetry along the main axis with very fast bipolar outflows.

Previous kinematic studies on Hb~5 have been conducted by \citet{HB5CR93} and \citet{HB5PP00}, they both have pointed out the large expansion velocities observed in this object, in the order of  
\vexp ~$\simeq$ 250 \kms, and the high electron densities in the core as compared to the lobes. 
\citet{HB5PP00} also mention detection of diffuse gas near the core reaching velocities near 400 \kms, however the corresponding evidence is not shown in their work and its origin has remained uncertain. WFPC 2  images of Hb~5 are available in the {\it HST} archive, these were obtained in 1998 under proposal ID 6502 (Balick, B., Icke, V. \& Mellema G). Large amounts of molecular hydrogen in the core of Hb~5 have been reported by \citet{Hb5P07}, they also confirm the Type I classification (He- and N-rich) for this nebula and conclude that Hb~5 has descended from a relatively massive progenitor star (M $>$ 4 M$_{\odot}$). For an uncertain distance of 3.2~kpc they derive an effective temperature for the central star \teff  ~= 170,000 K a stellar radius R = 0.17~R$_{\odot}$ and a luminosity L~=~2$\times$10$^4$~L$_{\odot}$. Previous statistical distances for Hb 5 have been estimated in the range 0.7~kpc -- 2~kpc \citep{hb5A92}. \citet{Hb5MR09} have reported the detection of X-ray emission apparently tracing the brightest optical features of Hb~5, namely the core and the point-symmetric filaments, they suggest that this emission may arise from shock-heated gas in the nebula.

There are no previous kinematic studies on K 3-17. Manchado et al.  (2000) present an image of K 3-17 but the spatial resolution is poor and key details are missing there. Good quality, ground based, narrow band imagery of K 3-17 have been published recently by  \citet{Miran10}, where the salient features can be clearly appreciated, particularly the nebular, rosette-type, core structure, and its  similarity with Hb 5 is noted. There are no images of  this nebula in the {\it HST} archive.

In this paper we aim now to  expand on previous results by inspecting the kinematics and morphology of these object in much finer detail using our extended spectroscopic coverage and the available high spatial resolution imagery. For Hb~5 the data are combined with a  morpho-kinematic model that gives us insights into the true 3D structure and expansion modes of this nebula. We present the first kinematic analysis of K 3--17, draw its similarities to Hb~5 and extend the conclusions of the modeling process for both objects. We also note that bipolar nebulae with rosette-type nuclei are likely precursors to poly-polar nebulae, such as NGC~2440 \citep{lop98}. We describe the observations
and data reduction in section \S2, discuss the morphology in section \S3 and the kinematics in section \S4; in section \S5 we describe the 3-D SHAPE model of Hb 5 and finally our conclusions are presented in Section \S6.

%\newpage 
\section{Observations and data reduction} 
\label{sec:observations}

High-resolution, long-slit, spectroscopic observations of Hb 5 and K
3-17 were obtained at the Observatorio Astron\'omico Nacional at San
Pedro M\'artir (SPM), Baja California, M\'exico, with the Manchester
Echelle Spectrometer (MES-SPM) \citep{Meaburn03} on the 2.1~m
telescope in its $f$/7.5 configuration.  For Hb~5 the observing runs
took place in 2003 June 7--9 and 2004 June 15. MES-SPM was equipped
with a SITE-3 CCD detector with 1024$\times$1024 square pixels, each
24 $\mu$m on a side ($\equiv$0.312 arcsec pixel$^{-1}$).  We used a 90
\AA~ bandwidth filter to isolate the 87$^{th}$ order containing the
H$\alpha$ and \nii{} $\lambda$$\lambda ~$6548,~6584~\AA, nebular
emission lines.  For Hb~5 we obtained twelve long-slit spectra stepped
across the nebula with the slit oriented north-south (i.e. position
angle P.A. = 0\degr{}). Two-by-two binning was employed in both the
spatial and spectral directions. Consequently, 512 increments, each
0\farcs624{} long gave a projected slit length of 5\farcm32 on the
sky. We used a slit 150 $\mu$m{} wide ($\equiv$~11~\kms{} and
1\farcs9).  Three additional spectra were obtained oriented along the
major axis of the lobes at P.A. $=$~80\degr. In this case we used a
slit 70~$\mu$m wide ($\equiv$~6~\kms{} and 0\arcsec.95) and no on-chip
binning.  K~3-17 was observed in 2010 August 26--27. MES-SPM was
equipped then with a Thompson CCD detector with 2048$\times$2048
square pixels, each 15 $\mu$m on a side ($\equiv$~0.195 arcsec
pixel$^{-1}$). A $3~\times~3$ binning was employed in this case for
682 increments, each 0\farcs585{} long that gave a projected slit
length of 6\farcm56 on the sky. We used a slit 150~$\mu$m{} wide
($\equiv$~11 \kms{} and 1\farcs9). Five slit positions were obtained
oriented along the mayor axis of the lobes at P.A. = $-$65\degr{}
(295\degr).  All the spectroscopic integrations for both, Hb~5 and
K~3-17, were of 1800~s duration. These spectra are part of the SPM kinematic
catalogue of galactic planetary nebulae \citep{lop12}, http://kincatpn.astrosen.unam.mx.

 Using MES-SPM in its imaging configuration we obtained a 60 s image
 for Hb~5 through the 90 \AA~wide H$\alpha$ + [N~II] filter and a deep
 1200~s image of the field of K~3-17 using a narrow (10~\AA{} HPBW)
 \nii $\lambda$ 6584~\AA{} filter. In Figure 1 are indicated the
 positions for the long slits on Hb~5 and K 3-17 shown against these
 SPM images, respectively. The seeing in the SPM images and spectra varied between 1\farcs0 and 1\farcs5 during the observations.

The spectra were calibrated in wavelength against the spectrum of a Th/Ar
arc lamp to an accuracy of $\pm$1 \kms{} when converted to radial
velocity.  The data were reduced using standard IRAF\footnote{IRAF is
  distributed by the National Optical Astronomy Observatories, which
  is operated by the Association of Universities for Research in
  Astronomy, Inc. under cooperative agreement with the National
  Science foundation} routines to correct bias, remove
cosmic rays and wavelength calibrate the two-dimensional spectra.

\section{Morphology} 

Narrow-band images of Hb~5 and K 3-17 are presented
in Figure 2. For Hb 5 the [N~II] {\it HST} image is shown on the left panel. 
The \NIIlam ~image of K 3-17 obtained at SPM is shown on the right panel of this Figure, but see also the superb images published by \citet{Miran10} on this object that show in much better detail its complex structure.
Both nebulae show a tight waist with a bright core. From the core emerge a complex set of filaments distributed as in a rosette configuration. For both, Hb 5 and K 3-17, some of these filamentary loops are proven for the first time in this work to actually be secondary expanding lobes. The primary bipolar lobes are also highly structured and filamentary,  displaying a density or brightness distribution along the major axis that gives a point-symmetric appearance to the overall bipolar shape. This can be clearly appreciated for Hb 5 in Figure 2, the same is true for K 3-17, though it is less conspicuous. In the latter, point-symmetry runs from a knot in the northern wall of the western  lobe to the southern wall of the eastern lobe. 
In both cases the lobes give the impression of being closed at the tips, however, the emission line profiles and their modeling (see next sections) indicate they actually end up open. 

\section{Kinematics} 
\label{Longslit spectra}

The calibrated position-velocity (P--V) representations of the
observed bi-dimensional line profiles corresponding to the \NIIlam
~emission from Hb~5 for slit positions a -- l (PA = 0\grados) are
shown in Figure 3 and for slit positions m -- o (PA = 80\grados) in
Figure 4. Spatial offsets are in arc seconds with respect to the
position of the central star. All spectra presented in this paper are
transformed to the heliocentric reference system (\vheli). The P--V
arrays are labeled according to their corresponding slit
positions. For each slit position we show a couple of P-V arrays, the
observed \NIIlam{} P-V array is on the left panel and the right panel
is the corresponding synthetic P-V array derived from the SHAPE
model. The spectra corresponding to slit positions a -- e in K 3-17
(PA = $ -65$\grados) are presented in Figure 5 in a similar way.

\subsection{Hb~5}

The heliocentric systemic velocity of Hb~5, \vsys $= -$17.5~\kms{},  has been derived from slit n that passes over the nebula's geometric center. 

From Figure 3 it is clear that the eastern lobe (slits a -- f) is blueshifted whereas the western lobe (slits h -- l) is redshifted with respect to \vsys. This effect can also be seen in Figure 4 as the tilt in the X-shaped line profiles that reflect the bipolar symmetry. The upper part of the line profile (eastern lobe) is blueshifted with respect to the middle of the P--V array, and the lower part (western lobe) is redshifted. From the information contained in the line profiles in Figure 4 and  the SHAPE model (see below) an inclination of the nebula of 28\grados $\pm$ 3\grados with respect to the plane of the sky is derived, in fair agreement with the previous estimates by \citet{HB5CR93} and \citet{HB5PP00}. 

Figure 3 shows the P--V arrays corresponding to the spectra that cut across the lobes. Since the lobes behave as expanding bubbles along the line of sight, the line profiles are close to velocity ellipsoids (VE). However, these VE show some unusual structure such as internal loops and knots and certain degree of deformation or lack of inner symmetry, all these elements contain information regarding the structure, geometry and velocity law in the lobes, as we shall discuss below and in the modeling section.  We shall mention that we found no trace in our spectra of the outer, faint ring-like shells that are visible in the equatorial region of Hb 5 in some {\it HST} images.

\subsection{Inner loops}
The clearest example of an internal loop in the structure of a VE is arrowed in the P--V array corresponding to slit position c in Figure 3, but it can also be seen in slit position d and less clearly in slit position e. These loops coincide in location with the filamentary point-symmetric structure of the east lobe, though presumably close to the back wall since they appear on the redshifted side of the VE. The loops are indicative of the presence of some sort of tube-like or long, warped inner structure in this region. 

The point-symmetric structure in the east lobe runs close to the upper edge but mainly along  the front wall of the lobe (see below) whereas in the western side it seems to run closer to the lower edge of the lobe since these internal loops are not apparent in the line profiles of the western lobe or simply there is not a closed structure along the filamentary bright region on the west side. 

\subsection{Deformed velocity ellipsoids}
In both cases, east and west lobes, the VE are deformed, tending in some cases to triangular shapes, see slit positions c -- f and their nearly mirror images i -- k in Figure 3. Slits a -- l are not perpendicular to the lobe axis and this tilt introduces some deformation in the observed profiles but this effect alone does not explain the shapes observed in the line profiles. Considering that the bi-dimensional line profile provides a representation of the structure of the expanding form along the line of sight, the presence of deformed VE indicate that  the shape of the lobes have evolved in a non-homologous way, most likely as a result of an expansion rate that contains in addition to a radial component from the core also a significant poloidal component. This effect is also clearly appreciated in the line profiles of Figure 4 where the line profiles show an apparent lack of symmetry in their bipolar shapes.

\subsection{Knots} 
Slit  a touches the very  tip of the blue lobe and
shows two high-velocity, low-ionization knots  at projected velocities, \vheli~$\simeq$~$-$200 and
$-$150~\kms. These knots are clearly overrunning the expansion of the eastern lobe which reaches its maximum blue velocity at that point at \vheli~$\simeq$~$-$120~\kms. Similar knots located close to the maximum blue velocity are found in the upper section of the VE for slits b -- f. 
The same situation is observed in the western lobe where the knots are found located in the lower section of the VE in slits i -- l, again close to the maximum red expansion velocities. These knots coincide spatially with the filamentary, bright, point-symmetric filaments and can be interpreted as tracing a distinct kinematic pattern for this structure. Their distribution along the point-symmetric regions of the lobes is also clearly shown in the line profiles from slits m \& o in Figure 4. The most straightforward interpretation in this case is to consider the set of filaments and knots that form the point-symmetric regions as outflowing material running along the inner walls of the opposite lobes. Considering again the inner loops located in the VE described above for the eastern lobe, it seems plausible to consider that this outflow has pushed away material from the wall of the lobe forming a distorted region. Close inspection of the {\it HST} image in Figure 2 shows a bay-like structure at the top of the eastern lobe (arrowed in that Figure) that coincides with the location of the inner loops detected in the line profiles of slits c \& d. Incidentally, this region coincides with the location of the extended X-ray emission detected in Hb 5 by  \citet{Hb5MR09}. Considering an angle with respect to the plane of the sky of 28\grados for the main axis of the nebula and assuming that the high-speed knots located outside and at the edge the VE that corresponds to the tip of the eastern lobe, observed at \vheli~$\simeq$~$-$200~\kms and $\simeq$~$-$150~\kms, respectively, represent the maximum radial velocity from the core of the outflowing gas along the lobe,  deprojected velocities \vexp $_{-rad}${} $\simeq$~$-$425~\kms{}  and \vexp $_{-rad}${} $\simeq$~$-$320~\kms{} are consequently derived. As for the transverse or poloidal lobe expansion, the VE  in Figure 3 and the line profiles in Figure 4 indicate that the maximum velocity separation across the lobes is  $\simeq$~220~\kms{} or a \vexp $_{-pol}${} $\simeq$~110~\kms{}  expansion rate, which in this case is overall little affected  by the inclination of the nebula with respect to the sightline. In these conditions it is reasonable to assume that the fast radial outflow from the core that produces the part of the point-symmetric structure in the east lobe interacts with the slower expanding wall, shock-heating gas in that region of the nebula and producing the extended X-ray emission, as suggested by  \citet{Hb5MR09}.  These radial and poloidal velocity components match well those derived from the SHAPE model, see section 5 below.

\subsection{Secondary lobes at the core}

The nebular nucleus of Hb~5 is very complex. A dense core is surrounded by large filamentary loops that resemble a rosette.  These filamentary loops seem to emerge from punctured regions of the core in random directions. If these filamentary loops indeed represent material escaping, likely due to a high pressure difference between the inner region of the core and its surroundings, the loops are then expected to actually be more like recently formed lobes. The analysis of these loops is complicated by their intricate relation with the rest of the nebula. There is however a large, well defined filamentary loop to the south of the core. Slit g crosses through it and a velocity loop can be discerned in its lower part, but it is slit o, see Figure 4, that confirms the lobe-like nature of it. This slit passes below the nebular core, cutting across the filamentary loop. The line profile shows a well defined velocity ellipse, nearly circular, at the center of the P--V diagram clearly indicating an expanding 3D structure. Likewise, in the case of slit m, in the upper part of the profile that corresponds to the eastern lobe, the red wall of the lobe emerging from the core in this line profile shows a nearly closed velocity loop corresponding in location to the filamentary loops located to the north-east of the core. \citet{HB5PP00} and \citet{Riera00} also noticed these expanding structures in their spectra,  calling them central blobs of gas and possible collimated outflows, respectively. These secondary small lobes when growing with time would be expected to produce a more defined poly-polar structure in Hb 5, likely similar to the case of NGC 2440 \citep{lop98}. 

It is also worth pointing out for slit n, Figure 4, that the line profile in this case is broad at the center of the profile due to the emergence of the secondary lobes and other complex density inhomogeneities in this region that we have not attempted to include in our  SHAPE model, thus the corresponding synthetic profile does not replicate this widening in the core.  
\subsection{K 3 - 17}

K 3 - 17 has remarkable similarities with Hb 5 (see Figure 2). A well
defined bipolar structure with a tight, dense waist, point-symmetric
arms and filamentary loops emerging from the core. The bi-dimensional
emission line spectra from slits a -- e for K~3-17, obtained along its
major axis, are shown in Figure 5. The heliocentric systemic velocity,
\vsys, is $+$9.2 \kms{} calculated using slit c. The spectra clearly
show the signature of faint bipolar lobes with large 
transverse expansion velocities, \vexp{}$\simeq$~100~\kms{}  along the 
sightline and a bright, dense
core.  On the western lobe slits b and c show bright, blueshifted
knots near the tip of the lobe. The eastern lobe is fainter and only
the emergence of the outflow close to the core is
detected. Interestingly, slits d and particularly e, clearly reveal
the presence of velocity ellipses at the location of the filamentary
loops located south of the core (see Figures 1 and 2) confirming again
their nature as secondary, equatorial lobes. Although our data for
K~3-17 are less detailed than for Hb 5 there is, nevertheless, little
doubt that they both show extraordinary akin characteristics and must have
had very similar conditions at the time of their formation.

\section{Morpho-kinematic modeling with SHAPE}
\label{sec:SHAPE}

The 3-D morpho-kinematic structure in the \NIIlam{} emission of Hb~5 has been modeled using the code SHAPE, developed by \citet{shapeWo06} and \citet{shapeWo10}. Similar analyses have been performed recently on the planetary nebulae NGC 6751 \citep{clark10} and NGC 6337 \citep{gd09}.
Modeling with SHAPE follows three main steps. First, defining the geometrical forms to use;
SHAPE has a variety of objects such as a sphere, torus, cone, cube, etc. whose basic forms can be  modified by the user (e.g. squeeze, twist, boolean, etc). Second, an emissivity distribution is assigned to each individual object or structure, and third, a
velocity law is chosen as a function of position. SHAPE gives as
result a two dimensional image and synthetic P-V arrays, which are
rendered from the 3D model to be compared with the observed data. The parameters of the model are then iteratively adjusted until a satisfactory solution is obtained.

Hb~5 was modeled using the {\it HST }image shown in Figure~2 and the P-V
spectra shown in Figures 3 and 4. Our model was built with a squeezed sphere 
to model the main and secondary lobes. To
model the high velocity knots, the emission was distributed throughout the volume of small spheres. Other structures were modeled as shells.  
A key characteristic of cylindrically symmetric and homologously expanding nebulae is that the P-V diagrams can be stretched in velocity in such a way that the outline of the image and the P-V diagram match. This match determines the factor of proportionality in the mapping between position and velocity. In the case of Hubble~5, the P-V diagrams cannot be stretched to match well to the outline of the image. Only a first order value of the factor of proportionality for an assumed linear relation between position and velocity can be estimated. There are three possible explanations for this mismatch. One, the expansion is homologous with all deviations from cylindrical symmetry being nearly along the line of sight. Second, the cross section is cylindrical and all deviations seen in the P-V diagrams are due to deviations of the velocity field from a homologous expansion. Third, a combination of the former two options. Option one is very unlikely for Hubble~5, since the deviations between image and P-V diagram are substantial and it seems extremely unlikely that such deviations from the roundish bipolar outline should be only along the line of sight. The basic cylindrically symmetric outline of the nebula as seen in images is therefore adopted and the corresponding velocity law for the main lobes was constrained by assuming this cylindrically symmetric cross-section, except for the presence of two bumps or distorted regions indicated by the line profiles.  This constrains the global velocity field to within an estimated 20 percent, including the poloidal component. This error estimate comes from trying a variety of combinations of structure and velocity fields. The error in the structure is therefore of similar magnitude. The general characteristics of the global velocity law in this case are similar to those found by \citet{SGSK09} in that the radial component increases monotonously faster than linear and there is a poloidal component that peaks at intermediate latitudes.
 For the main lobes and the high velocity knots we use a near quadratic  velocity law as a function of distance from the central star for the radial component, combined with a velocity law for the poloidal component (the component perpendicular to the radial vector) constraint by zero at the core and at the poles. Figure 6 shows the behavior of the components of the adopted velocity law for Hb 5. 
 We use a simple radial  Hubble law for the secondary bubbles. Figure~7 shows the resultant 3D mesh of the
model in several viewing angles before rendering. The top panel in this Figure shows the nebula as seen on the sky, the bottom left panel is a view of the nebula rotated 90\grados anticlockwise where the symmetric bumps in the opposite lobes can be appreciated and the last, bottom right, panel is a view of the nebula along the polar axis.

The results of the final rendered model are shown in figure~8, where
they are compared with the observations. Panel a) shows
the {\it HST} image of Hb~5, b) presents the observed \NIIlam{} profile from slit
 n (Figure 4), c) and d) are the corresponding rendered image and synthetic
P--V array from the slit position n.

\section{Discussion and conclusion} 
\label{sec:conclusion}

A thorough long-slit, spatially resolved, high resolution spectroscopic mapping of Hb 5 combined with {\it HST} imaging and a dedicated morpho-kinematic model have allowed a better understanding of the  structure of Hb 5 and by analogy also the closely similar bipolar nebula K 3-17, two singularly complex bipolar nebulae.
The high velocity, low ionization filaments and knots that produce much of the point-symmetric structure in Hb 5 are located on the inside, or close to the edge of the lobes, which are otherwise mostly hollow. The point-symmetric outflow is just reaching the tip of the lobes, which are open, and low ionization knots are starting to escape the lobes. 

In order to reproduce the line profiles a velocity law with radial and poloidal components is necessary, in this way the model reproduces with fair accuracy the most salient kinematic and morphological features of Hb 5, such as the deformed P -- V line profiles and the bumps in the lobes. The measured radial velocity component of the bipolar  outflow  reach \vexp $_{-rad}${} $\simeq$~$-$320~\kms{} at the tip of the lobe and we find a yet faster, isolated knot at  \vexp $_{-rad}${} $\simeq$~$-$425~\kms{}, probably part of the leading material from the fast, collimated outflow associated to the point-symmetric structures and now overruning the tip of the eastern lobe. The measured poloidal expansion of the lobes reaches  \vexp $_{-pol}${} $\simeq$~110~\kms{}.  Both, the radial and poloidal velocity components of the velocity law adopted in the SHAPE model match the observations.

The point-symmetric outflow denoted by low ionization filaments and knots is interacting with the walls of the lobes, this interaction produces shock-heated gas resulting in extended X ray emission previously detected by \citet{Hb5MR09}, particularly in the eastern lobe.

We find of particular interest the secondary lobes that emerge from the core of the nebula in a radial, rosette-type way, an extraordinary phenomenon that may lead to a complex poly-polar shape for Hb 5 in a near future as these secondary lobes develop and expand, ending in a case similar to NGC 2440 (L\'opez et al. 1998). This outstanding condition is shared by K 3-17. It is interesting to notice that the very young planetary nebulae Frosty Leo \citep{CC05} and AFGL 2688 \citep{Cox00} also display similar characteristics that may lead them to develop poly-polar and point-symmetric structures in the future, although it is difficult to predict the exact evolution of the different sets of lobes since additional mechanisms to pure ballistic motions will operate once the dense core grows and expands. Furthermore, in addition to the dynamical nebular evolution, illumination and ionization effects must also be considered for poly-polar nebulae, particularly in the younger ones, as discussed recently by \citet{Kwok10}.

There is little information on the nature of the central engine, beyond the inference by \citet{Hb5P07} of a single, relatively massive  (M $>$ 4 M$_{\odot}$) progenitor star. A  binary core might help explain the shaping and  point-symmetric outflows as originating from a precessing source, however no evidence of such a binary exists as yet. Point-symmetry as observed in Hb 5 can also be explained from single cores by the steady misalignment of a magnetic collimation axis with respect to the symmetry axis of the bipolar wind outflow, as described by \citet{GSL00}, but again, the observational evidence on the existence of a toroidal magnetic field is also lacking in this case. ALMA observations of the dense core environment of Hb 5 might reveal the necessary clues to understand the central engine of this PN.

 The picture that emerges from this analysis is that of a rapidly evolving ionized nebula from a compact core dominated by a dense equatorial mass distribution from which a fast bipolar outflow produces the main lobes, followed shortly after by an also fast but collimated bipolar outflow that now emerges tilted with respect to the main bipolar axis. This outflow interacts with the hollow lobes from the inside, continuing along and close to the the inner wall of the lobes, advancing at a similar speed to the radial expansion of the lobes. At a later time the dense, thick core starts to break up and secondary expanding lobes emerge from it in several directions, likely starting a pathway towards a poly-polar structure. 
 
 Considering a distance to the nebula of 1.7 kpc \citep{SH10}, a diameter (lobe length) of 27\farcs0 and the rate of radial expansion velocity, it is estimated that the development of the ionized structure, as seen today in Hb 5 has taken of the order of 1.5 $\times 10^3$ years. The density profile in Hb 5 is stronlgy peaked in the core and \citet{HB5CR93} estimate an average electron density of 1.5$\times 10^2$ cm$^{-3}$ for the lobes. Since the lobes are hollow, considering a width for the walls equivalent to 1\farcs0, the ionized mass contained in each lobe amounts to $\sim$ 3$\times 10^{30}$ grams or 0.0015 M$_{\odot}$. The seemingly episodic nature of  what appears to be different mass loss episodes may be due to actual discrete mass loss events from the central engine or to the varying conditions in the thick and dense circumstellar core environment which are reacting to the continuos erosion from mass loss and increasing photoionizing radiation pressure, bursting the core envelope in localized regions and producing fast expanding cavities through which stellar wind is channeled in different directions. Only a clearer understanding on the nature of the central engine will clarify this issue. 
  
\acknowledgments

This research has benefited from the financial support of DGAPA-UNAM
through grants IN116908, IN108506, IN100410, IN110011 and CONACYT 82066.  
We acknowledge the excellent support of the technical personnel at the OAN-SPM, particularly Gustavo Melgoza, Felipe Montalvo and Salvador Monrroy, who were the telescope 
operators during our observing runs. The authors thank the anonymous referee for his/her constructive comments that have helped to improve the presentation of this paper. The authors dedicate this work to the memory of their friend and colleague Yolanda G\'omez. 
\bibliographystyle{astroads}

\begin{figure*}
    \centering
  \includegraphics[width=1.0\textwidth]{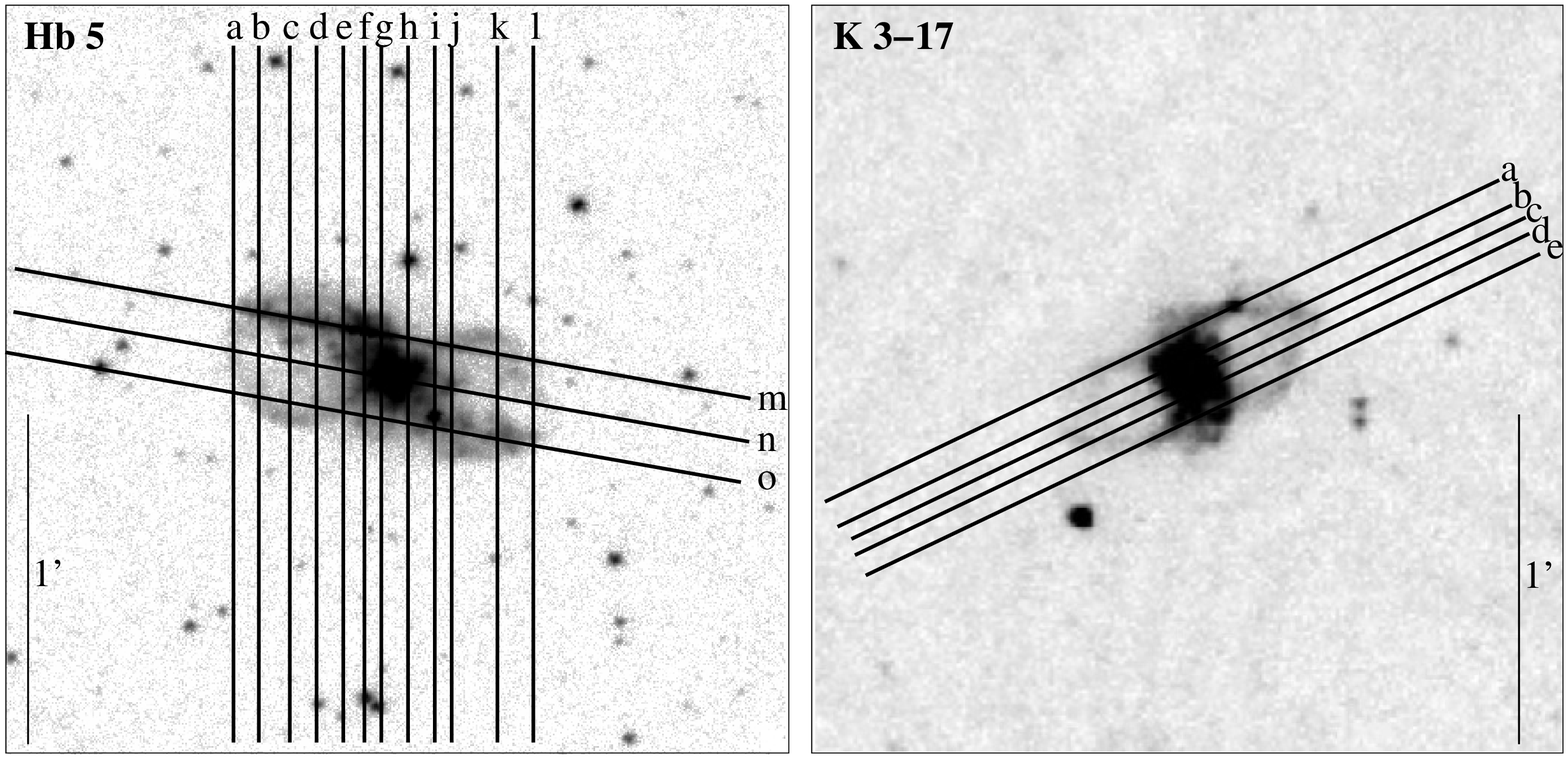}
  \caption{Location of each slit position is indicate and labeled on a
    H$\alpha$+\nii{} image from MES-SPM of Hb~5 and on a
    \nii{} image of K 3-17. North is up, east left in both cases.}
  \label{Fig.1}
\end{figure*}

\begin{figure*}
    \centering
  \includegraphics[width=1.0\textwidth]{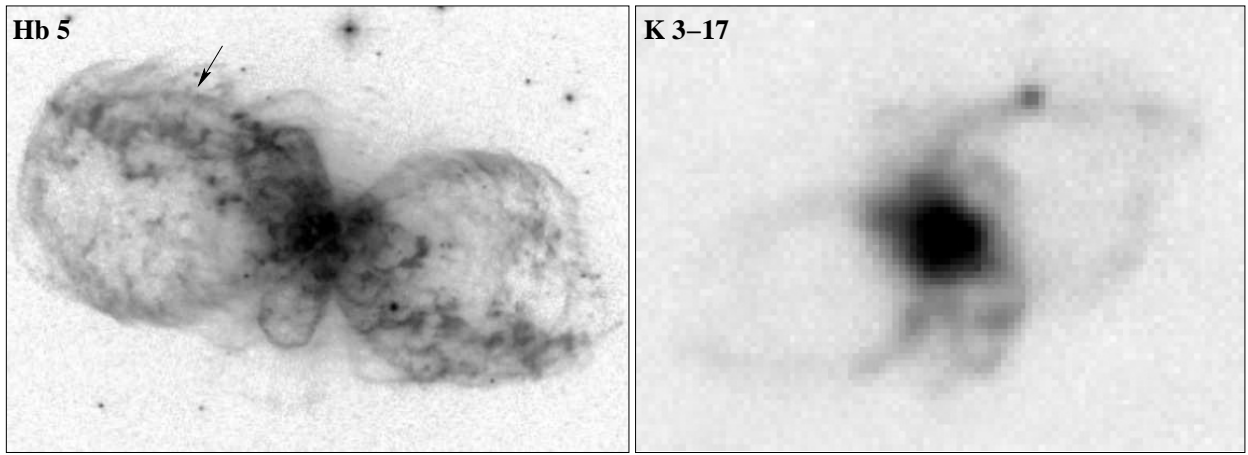}
  \caption{{\it Left Panel}: [N~II], {\it HST} WFPC2 image of Hb~5, the arrow indicates a bay-like feature discussed in the text. {\it Right Panel}:  [N~II], MES-SPM Image of K 3-17. }
  \label{Fig.2}
\end{figure*}

\begin{figure*}[!t] 
\centering
  \includegraphics[width=1.0\textwidth]{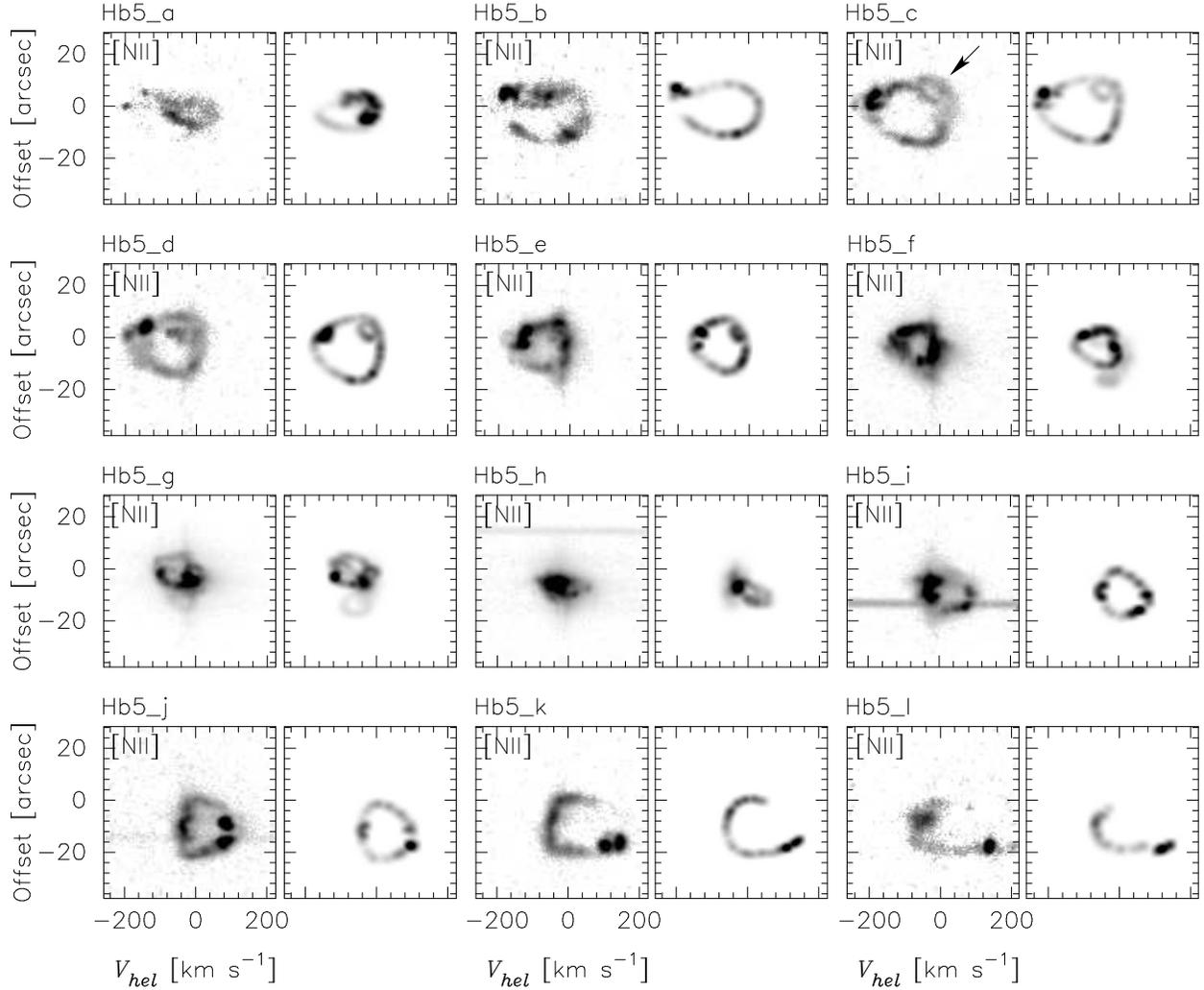}
  \caption{Mosaic of bi-dimensional (P -- V) arrays labeled
    according to slit positions of Hb~5. For each slit position we
    show a couple of P--V arrays, the observed \nii~6583~\AA{} P--V
    array is on the left panel and the corresponding synthetic P--V
    array derived from the model on the right}
    \label{Fig.3}
\end{figure*}

\begin{figure*}[!t]
  \centering
  \includegraphics[width=1.0\textwidth]{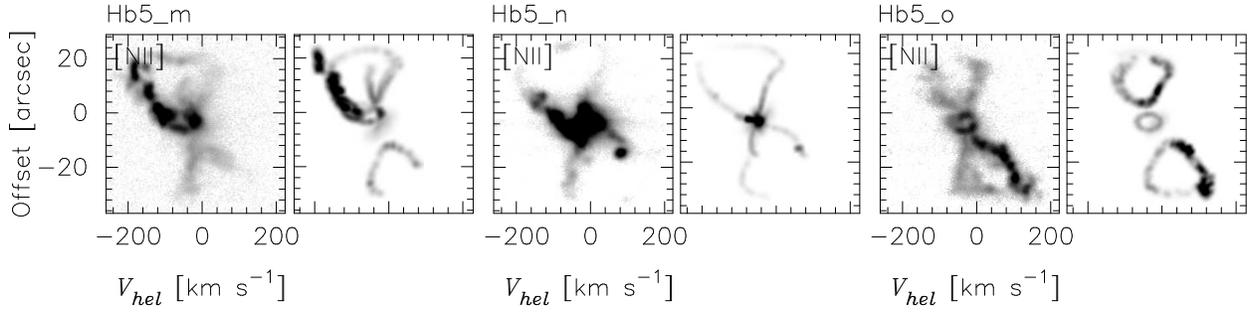}
  \caption{As  in Figure~3, but for slits
     m -- o in Hb~5, P.A. = 80\degr}
    \label{Fig.4}
\end{figure*}

\begin{figure*}[!t]
\begin{center} 
  \includegraphics[width=1.0\textwidth]{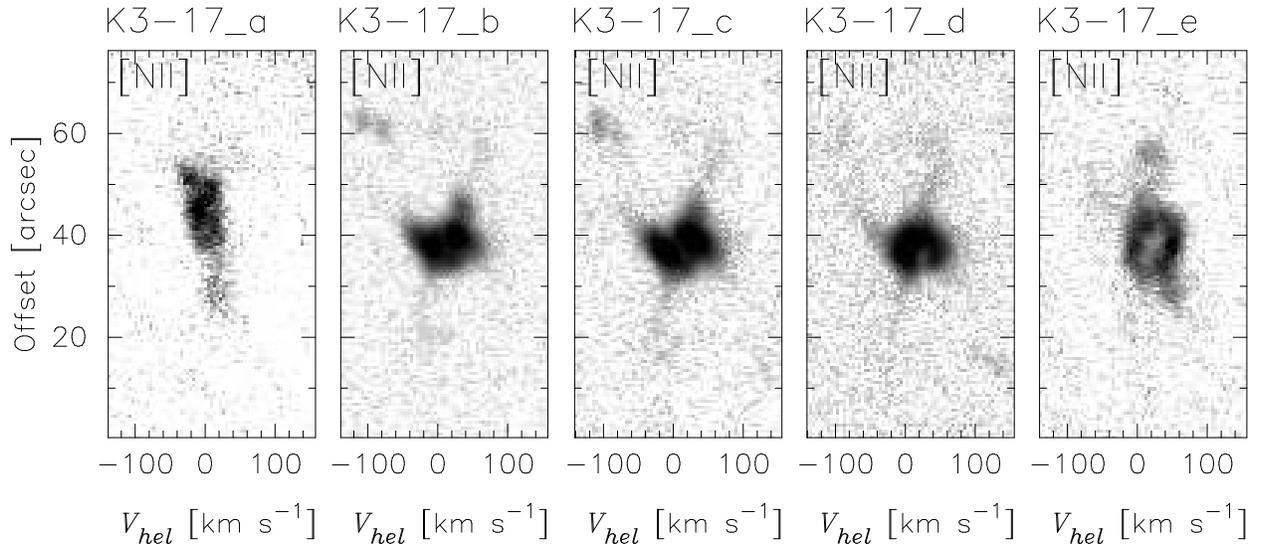}
  \caption{Mosaic of bi-dimensional (P -- V) of arrays of K 3-17 labeled
    according to slit positions . P.A. = $-$65\degr}
     \label{fig:spectra_k3-17}
\end{center}
\end{figure*}

\begin{figure*}[!t]
\begin{center} 
  \includegraphics[width=1.0\textwidth]{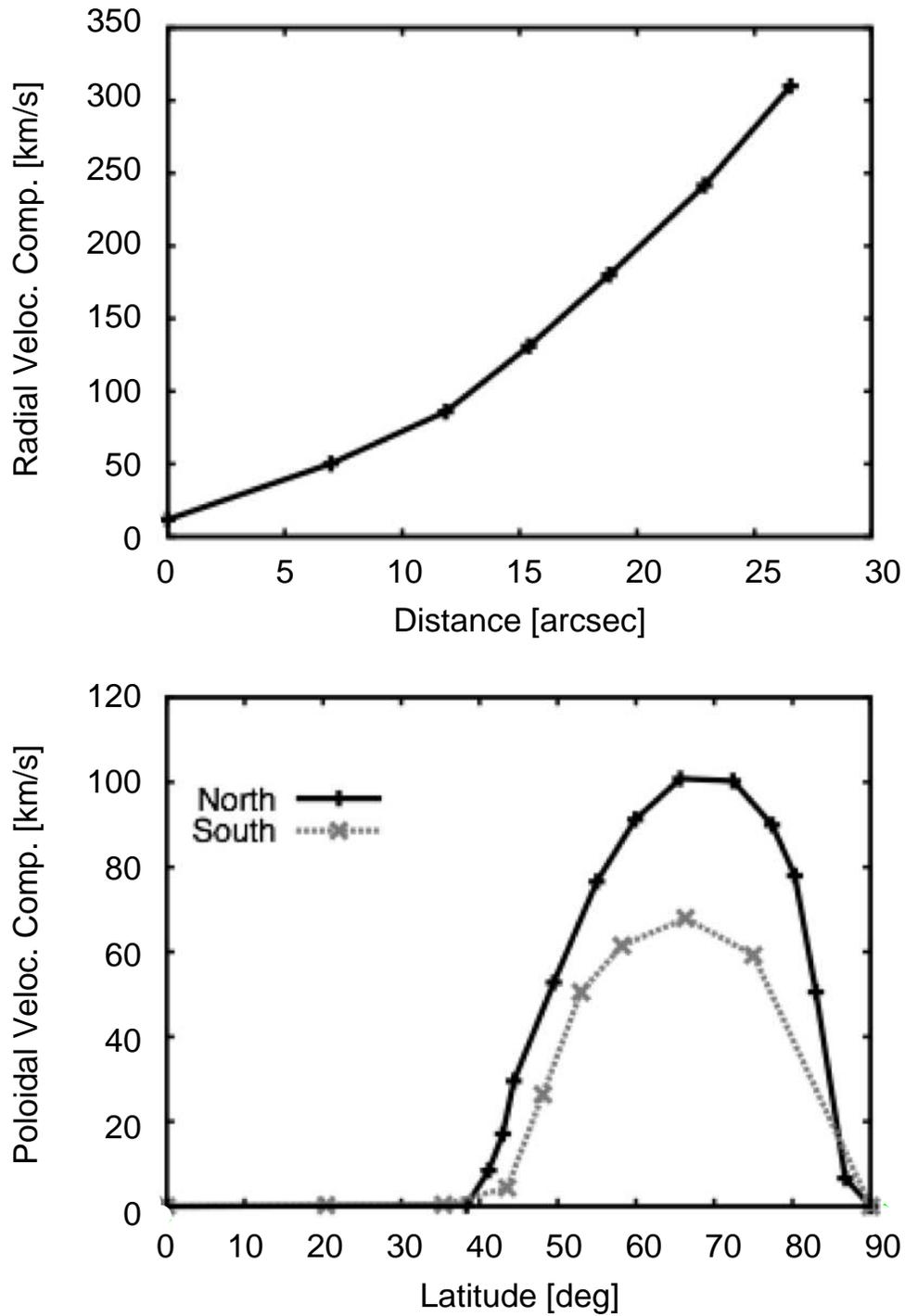}
  \caption{Top panel: The magnitude of the radial velocity component of the model is shown as a function of distance from the central star. The magnitude is the same for both lobes. Bottom panel: The magnitude of the poloidal velocity component is shown for the northern and southern lobes as a function of the angle form the symmetry plane (latitude). The position where non-zero values start is roughly the initial opening angle of the lobes.}
     \label{fig:velocity law}
\end{center}
\end{figure*}

\begin{figure}
\begin{center} 
  \includegraphics[width=1.0\textwidth]{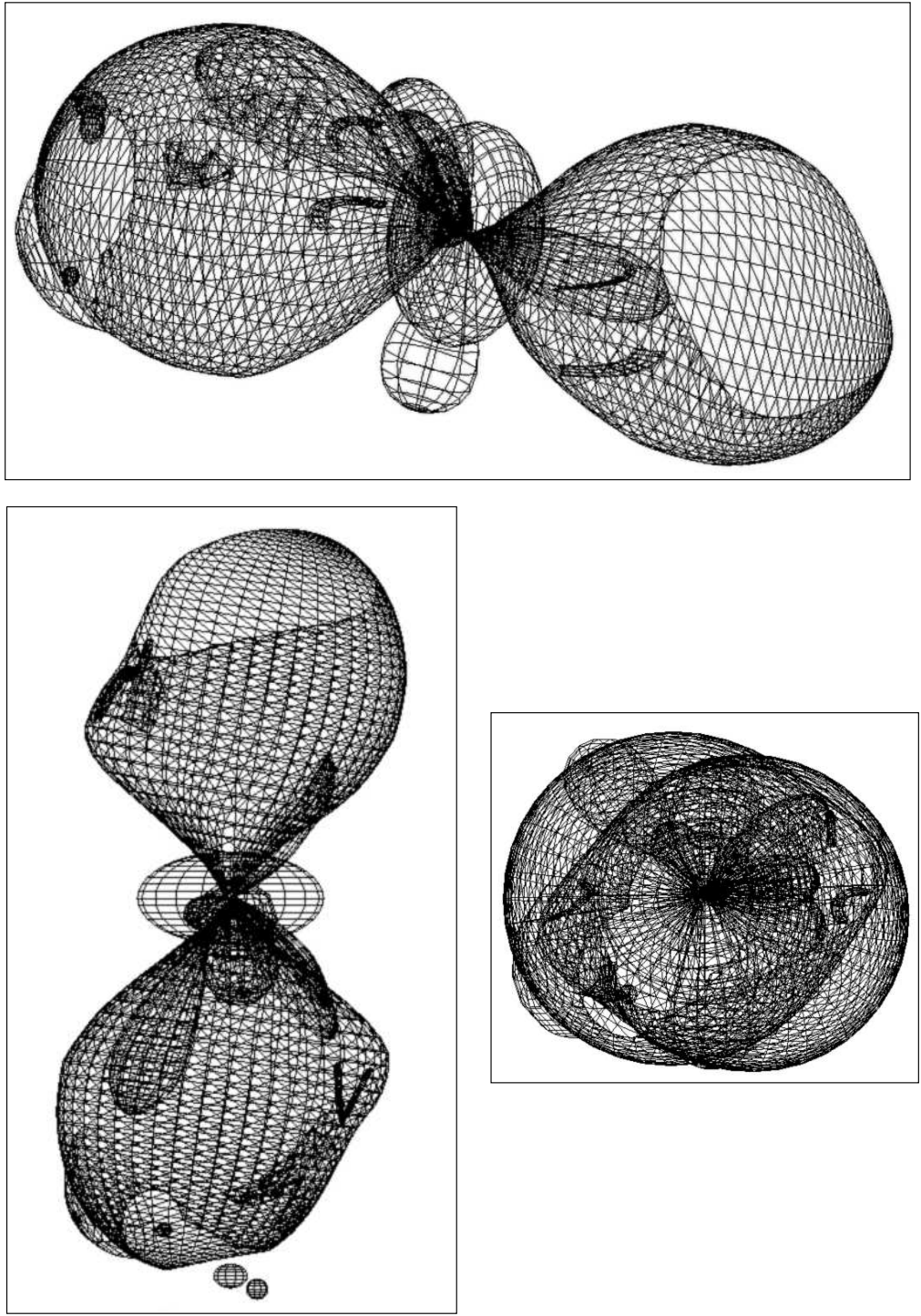}
  \caption{SHAPE mesh model of Hb~5 before rendering, shown at three
    different orientations. Top, as seen on the sky, bottom left, rotated 90\grados anticlockwise, bottom right, view along the polar axis.}
  \label{fig:shape}
\end{center}
\end{figure}

\begin{figure}
\begin{center} 
  \includegraphics[width=1.0\textwidth]{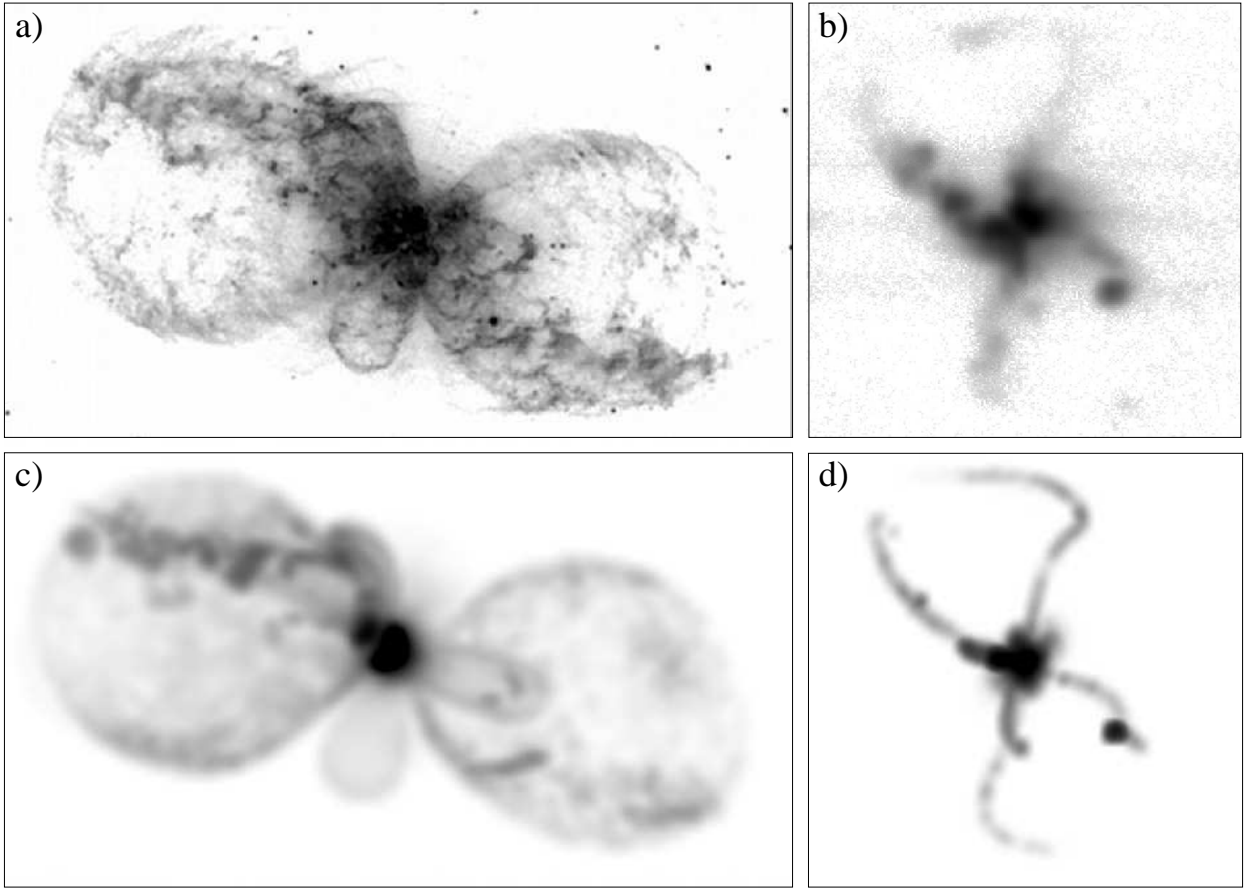}
  \caption{Top panels, frames a) and b), show the {\it HST} image of Hb 5
    and the observed \nii{} P-V array from slit n. Lower panels, c)
    and d) are the synthetic image and synthetic \nii{} P-V array from
    slit n, modeled with SHAPE}
  \label{Fig.7}
\end{center}
\end{figure}

\end{document}